\documentclass{ws-procs975x65}

\begin{document}

\title{THE BICONNECTION VARIATIONAL PRINCIPLE FOR GENERAL RELATIVITY}

\author{NICOLA TAMANINI}

\address{Department of Mathematics, University College London,\\
London, WC1E 6BT, United Kingdom\\
E-mail: n.tamanini.11@ucl.ac.uk}

\begin{abstract}
A recently proposed variational approach for general relativity where, in addition to the metric tensor, two independent affine connections enter the action as dynamical variables, is revised. Field equations always reduce to the Einstein field equations for any dependence of the matter action upon an independent connection.
\end{abstract}


\bodymatter

\section{Introduction}

Lagrangian formulations of general relativity abound in literature, but rather than differing in the choice of the Lagrangian density they differentiate one another by the variable upon which the variation of the action has to be taken. The most famous examples are pheraps the metric and Palatini variational approaches. The first one, sometimes called second order variation, considers the metric tensor as the only dynamical field variable while the second one, also known as first order variation, assumes the connection to be an independent field.

Though both approaches reproduce the Einstein field equations in vacuum, the latter gives rise to another theory whenever the independent connection is allowed to appear inside the matter action. In this case the theoretical framework goes under the name of metric-affine variational principle and it is known not to recover general relativity in the presence of spinor fields, for example.

The scope of the present paper is to summarize the basic features of a new variational approach recently proposed\cite{Tamanini:2012mi}. Together with the metric, two completely independent connections, with no {\em a priori} assumpions, are considered as dynamical variables entering the gravitational action for general relativity. The variation has thus to be taken independently with respect to the metric and both the connections. As we are going to show, applying this approach to a suitable generalization of the Einstein-Hilbert action leads to general relativity even when the matter action is allowed to depend on one of the connections. Because of the presence of two connections and in analogy with bimetric thoeries of gravity, this approach will be called {\em biconnection variational principle}.

\section{Action and Field Equations}

Denote with $\Gamma^{\lambda}_{\mu\nu}$ and $\Omega^{\lambda}_{\mu\nu}$ two independent connections, both containing torsion and non-metricity. In order to produce a new gravitational Lagrangian within our framework, we consider a generalization of the Riemann tensor obtained by symmetrizing the anticommutation of the covariant derivatives over the two connections. For any vector $V^\mu$ we have then
\begin{multline}
 \frac{1}{2} \left(\stackrel{\Omega}{\nabla}_\mu\stackrel{\Gamma}{\nabla}_\nu
      -\stackrel{\Gamma}{\nabla}_\nu\stackrel{\Omega}{\nabla}_\mu
      +\stackrel{\Gamma}{\nabla}_\mu\stackrel{\Omega}{\nabla}_\nu
      -\stackrel{\Omega}{\nabla}_\nu\stackrel{\Gamma}{\nabla}_\mu \right)V^\beta
 = \\
\left[ \frac{1}{2}{\Re_{\mu\nu\alpha}}^\beta(\Gamma,\Omega)
   +\delta_\alpha^\beta\left({S_{\mu\nu}}^\lambda(\Gamma) \stackrel{\Omega}{\nabla}_\lambda                             +{S_{\mu\nu}}^\lambda(\Omega)\stackrel{\Gamma}{\nabla}_\lambda \right) \right]V^\alpha \,,
\label{001}
\end{multline}
where ${S_{\mu\nu}}^\lambda(\Gamma)$, $\stackrel{\Gamma}{\nabla}$ and ${S_{\mu\nu}}^\lambda(\Omega)$, $\stackrel{\Omega}{\nabla}$ are the torsion tensors and covariant derivatives of $\Gamma$ and $\Omega$ respectively. The biconnection Riemann tensor ${\Re_{\mu\nu\alpha}}^\beta$ has been defined as
\begin{multline}
 {\Re_{\mu\nu\alpha}}^\beta(\Gamma,\Omega) \equiv \partial_\mu\Gamma_{\alpha\nu}^\beta
    -\partial_\nu\Omega_{\alpha\mu}^\beta +\Gamma_{\lambda\mu}^\beta\Omega_{\alpha\nu}^\lambda
    -\Gamma_{\lambda\nu}^\beta\Omega_{\alpha\mu}^\lambda \\
     +\partial_\mu\Omega_{\alpha\nu}^\beta
    -\partial_\nu\Gamma_{\alpha\mu}^\beta +\Omega_{\lambda\mu}^\beta\Gamma_{\alpha\nu}^\lambda 
    -\Omega_{\lambda\nu}^\beta\Gamma_{\alpha\mu}^\lambda \,.
\end{multline}
Notice that whenever the two connections coincide this reduces to the usual (metric-affine) Riemann tensor and (\ref{001}) becomes the standard formula for the commutation of covariants derivatives, which in general includes a torsion term.
The action we propose is thus
\begin{eqnarray}
 S_{BC}(g,\Gamma,\Omega) \equiv \frac{1}{4}\int d^4x\sqrt{-g}\,\, \Re \,,
\label{002}
\end{eqnarray}
where $\Re=g^{\mu\nu}\Re_{\mu\nu}=g^{\mu\nu}{\Re_{\mu\alpha\nu}}^\alpha$ in analogy with the Ricci tensor and the curvature scalar. When $\Gamma=\Omega$ this reduces to the Palatini (or metric-affine) action.

To (\ref{002}) we want to add a suitable matter action, where we have to make a choice upon the connection used to build covariant derivatives. In both metric and Palatini approaches this connection is assumed to be the Levi-Civita connection, while in metric-affine theories this is taken to be the independent connection which prevents the dynamical equivalence with general relativity. Since we will make the rather physical assumptions that all the matter fields couple to the same connection, in our case we have three choices: the Levi-Civita connection, $\Gamma$ or $\Omega$. The first choice would reduce the theory to Palatini general relativity and thus will not be considered. Because of the exchange symmetry $\Gamma\rightleftarrows\Omega$ of action (\ref{002}), it does not matter which one among this two connections enters the matter action. We choose $\Gamma$. The total action of the theory is thus
\begin{equation}
S= S_{BC}(g,\Gamma,\Omega)+S_{M}(g,\Gamma,\Psi)\,,
\label{005}
\end{equation}
where $\Psi$ denotes all matter fields collectively.

Variation with respect to $\Omega$ eventually yields the condition\footnote{The constraint $S_{\lambda\mu}{}^\lambda(\Gamma)=0$ has to be considered here because of a projective symmetry of (\ref{002}). See Ref.~\refcite{Tamanini:2012mi} for full details.}
\begin{equation}
\Gamma^\lambda_{\mu\nu} = \frac{1}{2} g^{\lambda\sigma} \left(\partial_\mu g_{\nu\sigma}+\partial_\nu g_{\mu\sigma} -\partial_\sigma g_{\mu\nu}\right) \,,
\label{003}
\end{equation}
which means that $\Gamma$ reduces to be the Levi-Civita connection. The role of $\Omega$ is thus similar to the one played by a Lagrange multiplier field. The variation with respect to $\Gamma$ leads, after one takes into account condition (\ref{003}), to
\begin{equation}
g^{\alpha\beta} {\mathcal{K}_{\beta\alpha}}^\nu \delta_\sigma^\mu + g^{\mu\nu}{\mathcal{K}_{\sigma\beta}}^\beta -g^{\lambda\nu} {\mathcal{K}_{\sigma\lambda}}^{\mu} -g^{\mu\lambda} {\mathcal{K}_{\lambda\sigma}}^{\nu} = 4\,\left( {\Delta_\sigma}^{\nu\mu}- \frac{2}{3}\Delta_\lambda{}^{\lambda[\mu} \delta^{\nu]}_\sigma\right) \,,
\label{004}
\end{equation}
where ${\mathcal{K}_{\mu\nu}}^\lambda = \Gamma_{\mu\nu}^\lambda-\Omega_{\mu\nu}^\lambda$ and ${\Delta_\sigma}^{\nu\mu}$ is the variation of the matter action with respect to $\Gamma$, the so-called {\em hypermomentum}.

Finally, variation of action (\ref{005}) with respect to the metric tensor produces, again after having taken into account (\ref{003}), the field equations
\begin{equation}
G_{\alpha\beta}+\frac{1}{2}\left(\delta_\alpha^\mu\delta_\beta^\nu -\frac{1}{2}g_{\alpha\beta}g^{\mu\nu}\right) \left({\nabla}_\lambda{\mathcal{K}_{\nu\mu}}^\lambda -{\nabla}_\mu{\mathcal{K}_{\nu\lambda}}^\lambda\right) =\bar{T}_{\alpha\beta} \,,
\label{006}
\end{equation}
where $\nabla$ is the standard covariant derivative of the Levi-Civita connection, $G_{\mu\nu}$ is the Einstein tensor and $\bar{T}_{\mu\nu}$ is the variation of the matter action with respect to the metric. Note that this differs from the usual energy-momentum tensor $T_{\mu\nu}$ of general relativity inasmuch as it lacks of the variation of the connection with respect to the metric. Using now equations (\ref{004}) and some algebra, we can rewrite field equations (\ref{006}) as
\begin{eqnarray}
G^{\alpha\beta} = \bar{T}^{\alpha\beta} +\nabla_\lambda \left(\Delta^{\beta(\alpha\lambda)}+\Delta^{\alpha(\lambda\beta)} -\Delta^{\lambda(\beta\alpha)}\right) \,.
\label{007}
\end{eqnarray}
The divergence in the right hand side of (\ref{007}) represents exactly the missing variation of the connection with respect to the metric in the tensor $\bar T_{\mu\nu}$\cite{Tamanini:2012mi,Hehl:1978}. We thus have that (\ref{007}) simply reduces to
\begin{align}
G_{\mu\nu} = T_{\mu\nu} \,,
\end{align}
which are nothing but the {\em Einstein field equations}.

\section{Conclusions}

To conclude, the biconnection variational approach for general relativity we have presented here, is physically equivalent to the more standard metric formulation of the theory. It allows the connection employed to minimally couple matter fields to be completely independent of the metric. In this respect, the biconnection approach is similar to the metric-affine one but, in contrast with the latter one, it always permits to reduce the equations of motion to the Einstein field equations and thus do not alter the phenomenology of gravitation. Applications of the biconnection formulation of general relativity can be relevant in quantum gravity theories, such as Loop Quantum Gravity, which assume a metric-affine classical framework.

\end{document}